\documentclass[utf8,a4paper]{article}
\usepackage{url,microtype,subcaption}
\usepackage[hidelinks]{hyperref}
\usepackage[onehalfspacing]{setspace}
\usepackage{comment}
\usepackage[numbers,sort&compress]{natbib} 
\usepackage{graphicx}
\usepackage{array}
\usepackage{float}
\usepackage{cite}
\usepackage{xcolor}
\usepackage{indentfirst}
\usepackage{array}
\usepackage{makecell}
\usepackage{multirow}
\usepackage[title]{appendix} 
\usepackage{cleveref} 
	
\usepackage{lineno}

\newcolumntype{C}[1]{>{\centering\arraybackslash}p{#1}}

\newcommand{\revision}{} 
\newcommand{\revisionb}{} 
\newcommand{\revisionc}{} 
\newcommand{\revisiond}{}

\begin{document} 
\title{Artifact detection and localization in single-channel mobile EEG for sleep research using deep learning and attention mechanisms}

\author{Khrystyna Semkiv$^{\#,a}$, Jia Zhang$^{\#,b}$, Maria Laura Ferster$^{b}$, and Walter Karlen$^{a,b}$\\
\small$^{a}$Institute of Biomedical Engineering, Ulm University, Albert-Einstein-Allee 45, 89081 Ulm, Germany.\\ \small e-mail: walter.karlen@ieee.org;\\
\small $^{b}$Mobile Health Systems Lab, Department of Health Sciences and Technology, ETH Zurich, Switzerland.\\ 
\small $^{\#}$equal contribution;
}
\def\corrAuthor{Walter Karlen, Institute of Biomedical Engineering, Ulm University, Albert-Einstein Allee 45,  89081 Ulm, Germany}
\def\corrEmail{walter.karlen@ieee.org}
\maketitle

\begin{abstract} 
Artifacts in the electroencephalogram (EEG) degrade signal quality and impact the analysis of brain activity. 
{\revisiond Current methods for detecting artifacts in sleep EEG range from threshold-based algorithms to machine learning approaches, yet applications remain limited for single-channel mobile EEG.}
We propose a convolutional neural network (CNN) model incorporating a convolutional block attention module (CNN-CBAM) to detect and localize artifacts in sleep EEG using attention maps.  We benchmarked this model against 6 other machine learning and signal processing approaches. We trained/tuned all models on 72 manually annotated EEG recordings obtained during home-based monitoring from 18 healthy participants with a mean (SD) age of 68.05 y ($\pm$5.02). We tested them on 26 separate recordings from 6 healthy participants with a mean (SD) age of 68.33 y ($\pm$4.08), which contained artifacts in 4\% of epochs. CNN-CBAM achieved the highest area under the receiver operating characteristic curve (0.88), sensitivity (0.81), and specificity (0.86) among the tested approaches.  Under the ideal choice of an attention threshold of 0.66, the attention maps from CNN-CBAM  localized artifacts within detected artifact epochs with a sensitivity of 0.61 and specificity of 0.63.
This work demonstrates the feasibility of automating artifact detection and localization in wearable sleep EEG.

\end{abstract}

\section{Introduction}

Electroencephalography (EEG) is a direct and non-invasive measurement of electrical brain activity and is an essential modality for brain research \citep{Raduntz2018, Casson2019} and clinical diagnosis. EEG monitoring plays a critical role for studying sleep \citep{Patanaik2018}, human cognition \citep{Zarjam2015}, mental states \citep{Zeng2018}, brain-computer interfaces \citep{Lebedev2017, Casson2019}, as well as various health conditions \citep{Raduntz2018}. In many of these applications, a high-density scalp recording system with many EEG channels is used. However, such a recording setup is bound to a laboratory setting and requires tedious preparation steps with expert supervision. The EEG montage is usually uncomfortable and is accompanied by a decline in signal quality over time. Therefore, it is unsuitable for long-term monitoring or remote applications without expert supervision. Consequently, there is an emerging interest in EEG technologies that enable remote monitoring without time constraints and geographical limitations. Wearable EEG devices have been developed to provide users with a comfortable, user-friendly way to self-monitor EEG in the wild with minimal disruption to everyday life. These devices target research applications, such as drowsiness detection \citep{Li2015}, sleep-wake monitoring \citep{Sterr2018}, or slow-wave modulation with auditory stimulation during sleep \citep{Ferster2019}, but also clinical applications in a remote setting can be envisioned, such as long-term screening for epilepsy or sleep disorders. {\revisionb Many of these devices are Internet of Medical Things (IoMT) ready. IoMT-enabled wearables are embedded within a larger healthcare infrastructure, enabling quasi-real-time data streaming and cloud-based EEG processing and analytics.}

{\revisionb An inherent challenge of both laboratory and ambulant EEG is noise and various types of artifacts that can corrupt the signal. Signal quality loss can have technical sources arising from interactions between devices and the environment, or physiological sources, such as interferences from biosignals other than brain activity. Technical artifacts are often linked to electrode materials and placement, as well as cables. For example, electrode displacement during subject movement generates motion artifacts \citep{Craik2019}. Altering the pressure on the electrodes influence the contact impedance, introducing slow DC variations \citep{Nathan2016}.
Novel electrode materials \citep{Leach2020} and a changing environment can also induce noise \citep{Teplan2002}.
Typical physiological,  high-amplitude artifacts are eye movements and blinking. Sweating and the activity of sweat glands lead to low-frequency variations. Alternating muscle activity results in amplitude variations and high-frequency noise. Even the electric activity of the heart can cause rhythmic spikes visible in the EEG.}

Inadequate filtering can mask the aforementioned noise contamination and, consequently, lead to false task-related interpretations. For example, scalp EEG contaminated with eye movement artifacts can prevent effective real-time brain-machine interfaces \citep{Kilicarslan2016}. Unsurprisingly, artifact detection and removal are standard routines in EEG analysis, {\revisionb often conducted manually \citep{dsn, Leach2020, Craik2019, Tsinalis2016}.} The manual identification is time-consuming, labor-intensive, and tedious. For large-scale deployment of wearable, distributed EEG monitors, visual inspection of the EEG becomes unfeasible due to the massive volume of data generated. Therefore, it is essential to find artifact identification approaches that can automatically identify drops in signal quality in continuous EEG data streams {\revisionb received from remote wearable EEG devices}.  

Several {\revisionb signal-processing} approaches have been proposed to automate EEG artifact detection. A straightforward but somewhat limited approach is to set a threshold on the EEG amplitude {\revisionb in the time-domain} or the signal-to-noise ratio in the frequency domain \citep{Raduntz2018}. Other approaches target specific EEG artifacts, such as removal of ocular artifacts (eye movements and blinks) with an independent component analysis and wavelet transform \citep{Burger2015}, setting thresholds based on a statistical distribution of standard deviations of EEG amplitudes in different sleep stages \citep{Rozario2015}, or motion artifact detection using external gyroscope sensors \citep{Regan2013}. In the context of sleep research, the goal is to exclude EEG epochs corrupted by artifacts, which would impact automated sleep staging \citep{Malafeev2019}. However, these methods have been mainly designed and evaluated for laboratory settings where high-density EEG recording systems are used, and conditions are controlled. The type and occurrence of artifacts significantly alter when the EEG originates from wearable devices used in the wild. Therefore, deploying efficient and efficacious algorithms that can automatically process large quantities of EEG to identify artifacts would be desirable. They could inform end-users or researchers in {\revisionb quasi real-time} about quality drops or prevent inadequate interventions delivered with the EEG device.   
  
Our goal was to develop and validate an automated method to detect and localize EEG artifacts, specifically to {\revisionb cloud-based analysis} of long-term sleep data from IoMT-enabled wearables. In this work, we focused on designing a deep learning model to accomplish a binary detection of artifacts in {\revisionb raw} sleep EEG  obtained from remote home monitoring. {\revisionb In addition, we integrated a machine learning attention mechanism to facilitate the easy localization, visualization, and interpretation of artifacts within a streamed EEG epoch. Our approach offers a scalable, interpretable solution for continuous remote sleep EEG monitoring, enabling artifact detection with minimal preprocessing.}

\subsection{Related work} 
Traditionally, detecting artifacts in EEG has relied on signal processing and extensive hand-crafted feature extraction methods \citep{Raduntz2018, Burger2015, Rozario2015}. {\revisionc In the past decade numerous reviews have analyzed and compared these approaches      \citep{islam2016,sadiya2021,mannan2018}
}. To summarize, these approaches are time-consuming and require multistep preprocessing of sensor data, which impedes transfer to different hardware and setups \citep{Bahador2020}.
With advances in deep learning and increased data availability, artificial neural networks have been extensively explored for analyzing physiological signals and time series, including EEG. 
Deep learning architectures based on convolutional neural networks (CNNs), recurrent neural networks (RNNs), and transformers can learn features directly from raw data with minimal preprocessing and feature engineering, {\revisionb making them more powerful than traditional machine learning methods.  For instance, transformer-based architectures have been introduced to match the noisy input signals to their counterparts \citep{Chuang2025, Cai2025}. The models were learning to generate a clean output signal from a single noisy input based on paired data consisting of noisy signals and their denoised versions. Although these models are primarily designed for artifact removal, some incorporate artifact detection as an intermediate step to better isolate contaminated EEG segments before correction. For example, Saba-Sadiya et al. classified artifact segments prior to applying EEG correction based on outlier detection \citep{saba2021}.  Peh et al. developed a CNN-based cascade model integrated in a transformer to detect five types of artifacts on the channel level and determine whether a multi-channel EEG segment was artifact-free or not \citep{peh2022}. Once identified, the corrupted segments were interpolated using data from adjacent clean segments. %
Aquilu\'e-Llorens et al. proposed a long short-term memory (LSTM)-based autoencoder architecture, leveraging the reconstruction error between input and output signals to identify artifact segments \citep{lorens2025}.
Van-Stight et al. introduced a shallow CNN model for artifact detection. However, instead of reconstructing it, the contaminated segments are rejected \citep{Stigt2023}. 
Although all these strategies vary in design, they operate on the entire segment.  
This means that if an artifact is partially present in a segment, the non-artifact regions are corrected or discarded as well. This introduces potential inaccuracies in the reconstructed signal or, in the case of artifact rejection, unnecessary discarding of clean data, leading to a reduction in dataset size. This can be especially problematic for biomedical tasks where the data availability is limited.

An alternative would be to localize the exact temporal boundaries of the artifacts. This would allow models to operate more effectively, avoiding unnecessary corrections or data loss. Yu et al. and Seeuws et al. proposed deep learning methods that identify the center point and duration of the artifacts \citep{Yu2024, Seeuws2024}, which are referred to as point-level methods. Yu et al. show that with a learnable wavelet method (WaveNet) and a tree structure, artifacts can be localized and the contributing frequencies identified \citep{Yu2024}. Another approach used in deep learning, which can be applied directly in the time domain to localize individual artifacts in the signal, are attention mechanisms.  They have been successfully deployed in CNN models for discriminative feature representation \citep{Wang2017, Hu2018}, but have not been explored for identifying artifacts.}

\begin{figure}
    \centering
    \includegraphics[width=1\linewidth]{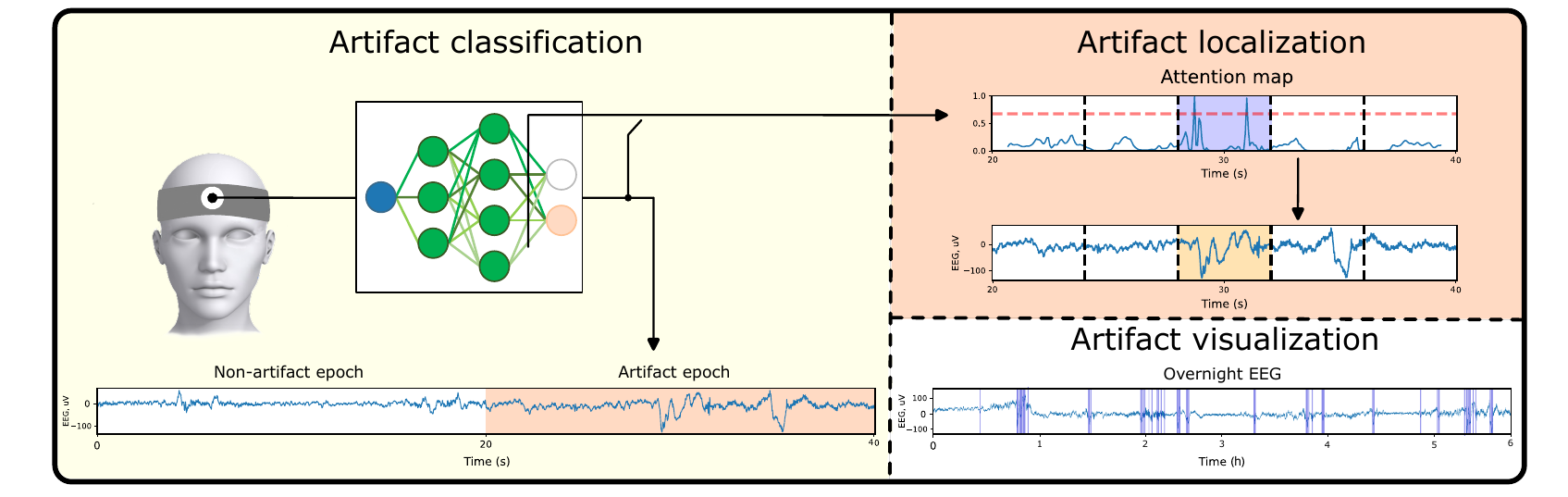}
    \caption{Artifact detection and localization from raw sleep EEG using a convolutional neural network model with an attention mechanism.}
    \label{fig:Figure_1}
\end{figure}

We propose a CNN-based model with an attention mechanism to detect and localize artifacts in sleep EEG signals (Figure \ref{fig:Figure_1}). 
Our contributions are: 

\begin{enumerate}
\item Development of a CNN-based end-to-end deep learning method to classify artifacts from raw EEG; 
\item Modification of {\revisionb the convolutional block attention module for EEG data streams and time-series};
\item Demonstration of the advantage of a CNN-based model with an attention mechanism over other CNN-based deep learning models and standard signal processing approaches;
\item A visualization strategy using an attention map to rapidly and accurately localize signal quality drop in streaming EEG data.
\end{enumerate}

With such solutions, sleep researchers can rapidly identify noisy regions in IoMT EEG recordings, generate data availability graphs, apply correction mechanisms, and prepare the data for further detailed analysis. 

\section{Methods and Materials}
{\revisiond The objective of this study was to develop and validate a deep learning framework capable of both detecting and localizing artifacts in single-channel EEG data collected from wearable devices in unsupervised home environments. Our methodological approach followed a three-stage pipeline:
\begin{enumerate}
\item \textbf{Model Development:} We designed four deep learning architectures with increasing complexity, starting from a baseline two-branch CNN and incrementally adding LSTM layers for temporal modeling and a 1D Convolutional Block Attention Module (CBAM) for feature recalibration. This incremental design allowed us to quantify the performance gains provided by each specific architectural component.
\item \textbf{Benchmarking:} To establish a rigorous performance baseline, we implemented three established open-source artifact detection methods representing different signal processing paradigms: spectral power analysis (frequency-domain), standard deviation thresholding (amplitude-domain), and a heuristic 1D-CNN (machine learning).
\item \textbf{Evaluation and Localization:} Beyond binary epoch classification, we utilized the attention maps generated by the CBAM layers to perform artifact localization. These maps were validated against expert-labeled 4-second windows to determine the model's ability to identify the precise temporal location of noise within a 20-second epoch.
\end{enumerate}

The following sections detail the dataset characteristics, the mathematical formulation of the proposed architectures, and the experimental setup used for benchmarking and validation.
} 

\subsection{Wearable EEG Dataset}
\label{dataset}
To train the models and compare artifact detection performance, we used a single-channel EEG dataset from a clinical trial (NCT03420677) \citep{Lustenberger2022}, in which healthy older adults wore a mobile device at home for multiple nights. The data originated from 24 Caucasian participants (10 female and 14 male) with a mean (SD) age of 68.12 y ($\pm4.72$). Healthy was defined as good general health, non-smokers, and no presence or history of a psychiatric/neurologic disorder, no diagnosed sleep disorder, or no internal disorder. Further details on the study design can be found in \citep{Lustenberger2022}. A total of 98 recordings with a median duration of 7.9 h (range 5.5 to 9.9 h) were available. The participants wore the Mobile Health Systems Lab Sleep Band (MHSL-SleepBand v2, \citep{Ferster2019}) that sampled biosignals at 250 Hz. Subjects self-applied electrodes at the central forehead (Fpz) and both mastoid positions for unipolar EEG derivation and common mode rejection. Electrooculogram (EOG) and chin electromyogram (EMG) were also derived, but not used in this study. The raw biosignals were recorded to an SD card for later analysis. 

{\revisionc Recordings were randomly distributed across four experts for sleep scoring. For each recording, one to three experts were assigned to perform manual scoring, resulting in most recordings having annotations from a random single scorer.  For the 12.5\% of recordings scored by multiple experts, a Cohen’s Kappa score of 0.57 was obtained. } Each night was scored for sleep stage epoch-by-epoch, using 20-second windows. Each expert also labeled artifacts on a 4-second basis during NREM and REM episodes, meaning that five 4-second windows within one epoch could be marked as an artifact. {\revisionb The following rules were considered to detect an artifact: 1) muscle/movement artifacts in the EEG when also seen in the EMG, 2) clear EEG large amplitude artifacts/spikes/very noisy signal (phasic). ECG artifacts were not considered, as they usually affect the whole recording.} Eye movement artifacts in the REM stage and sweating artifacts, which were seen as low-frequency sine wave-like amplitudes below 1 Hz, were also not considered artifacts. If at least one 4-second window was termed an artifact within a 20-second epoch, the entire epoch obtained an artifact label. 

\subsubsection{Data preprocessing}

\begin{figure} [h]
    \centering
    \includegraphics[scale=0.6]{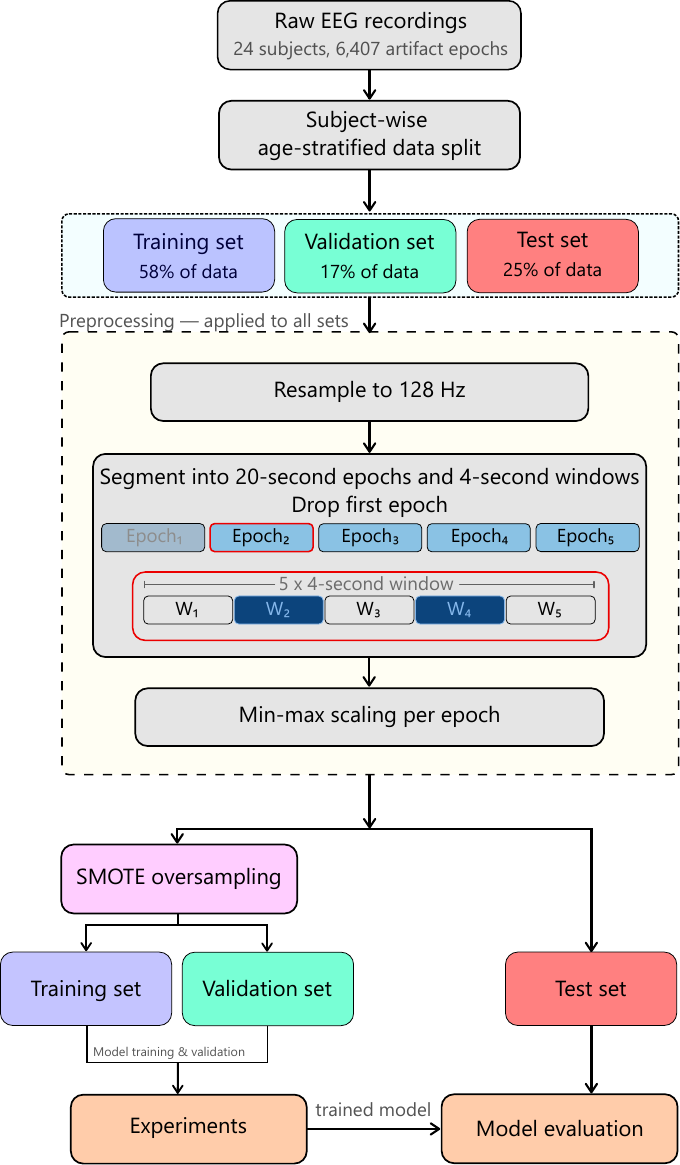}
    \caption{Data processing pipeline for the artifact classification model development and evaluation. Wearable, single-channel EEG data is split in training, validation and test set. After preprocessing (yellow box), artifacts in training and validation are augmented with SMOTE and used in the experiments to train the models. }
    \label{fig:Figure_data}
\end{figure}

As sleep EEG features vary across age groups, we split the data from 24 subjects by age and number of recordings per subject into training (58\%), validation (17\%), and test (25\%) sets (Table \ref{table:1}). {\revisiond We then resampled all raw EEG signals to 128 Hz and segmented them into 20-second non-overlapping epochs, each containing five 4-second windows. We clipped the first 20 seconds of each recording due to extreme values, likely caused by adjustment of the EEG band, movement artifacts, or noise. Subsequently, we applied min-max scaling {\revisionc (mapping to [0, 1])} for each 20s-epoch separately. 
Due to the heavy imbalance (approx. 5\% artifacts), we applied the Synthetic Minority Oversampling Technique (SMOTE) as the last preprocessing step to the training and validation sets \citep{smote2002} to maintain class balance during training and parameter tuning (Figure \ref{fig:Figure_data}). To validate that SMOTE generated plausible artifact epochs, we conducted a qualitative comparison between original and synthetic artifact samples by selecting pairs with matching spectral characteristics using power spectral density (PSD) analysis. The visual inspection confirmed that synthetic samples preserved the general morphological properties of real artifact epochs (Appendix \ref{appendix:A}).} 

\begin{table}[!htb]
\caption{\small Distribution of recordings, epochs, and artifacts across a mobile EEG dataset's training, validation, and testing sets.}
\centering
\setlength{\tabcolsep}{+4pt}
\setlength{\extrarowheight}{3pt}
\begin{tabular}{p{50pt}|c c c c c c}
\hline
& \thead{\small Participants \\ \small n. (\%)} 
& \thead{\small Age, mean \\ \small ($\pm$ SD)} 
& \thead{\small Recordings \\ \small n.} 
& \thead{\small EEG \\ \small segments, n.}
& \thead{\small Artifact \\ \small segments, n. (\%)} 
& \thead{\small Artifact segments \\ \small after SMOTE, n.} \\ [4pt]
\hline
\small Total       & 24           & 68.12 ($\pm$4.72)& 98 & & &\\
\small Training    & 14 (58.0 \%) & 67.92 ($\pm$4.98) & 57 & 81,619 & 3,888 (5.0 \%) & 77,731 \\
\small Validation  & 4 (17.0 \%) & 68.50 ($\pm$5.91) & 15 & 20,032 & 1,019 (5.3 \%) & 19,013 \\
\small Test        & 6 (25.0 \%) & 68.33($\pm$4.08) & 26 & 37,111 & 1,500 (4.0 \%) &\\ 
\hline
\end{tabular}
\label{table:1}
\end{table}

\subsection{Deep Learning Architecture Designs}
We composed all our deep learning models based on a CNN architecture, {\revisionb as CNNs efficiently capture local temporal and spatial patterns}. First, we developed a baseline two-branch CNN model, and then we complemented this model with an LSTM. In addition, we augmented both of these models with an attention mechanism. {\revisionb This incremental approach enabled {\revisiond an ablation study where the contribution of each architecture component to performance could be derived.}}

\subsubsection{Baseline two-branch CNN}
The two-branch CNN model was inspired by a model proposed for sleep stage classification \citep{dsn}. It consists of two separate branches, each featuring CNN layers with small and large kernel sizes, respectively {\revisionc (Figure \ref{fig:Figure_2}.a)}. The small kernel size focuses on a short time scale to capture time-domain information. In contrast, the large kernel size focuses on the expanded time scale, detecting repetitive patterns and frequency information in the EEG epochs. Both branches include five convolutional layers, each followed by batch normalization and, in some cases, by a dropout regularization mechanism and/or max-pooling operation. The outputs of the two branches are combined and passed through a fully connected layer to produce the final artifact classification prediction.
We used an EEG input vector of length 2560, corresponding to a single 20-second epoch for each CNN branch in our model. To adapt the model to our specific task, we modified the kernel sizes of the original design {\revisiond which can be derived from  Figure \ref{fig:Figure_2}a}.

\subsubsection{CNN-LSTM}
\begin{figure}
    \centering
    \includegraphics[width=0.75\linewidth]{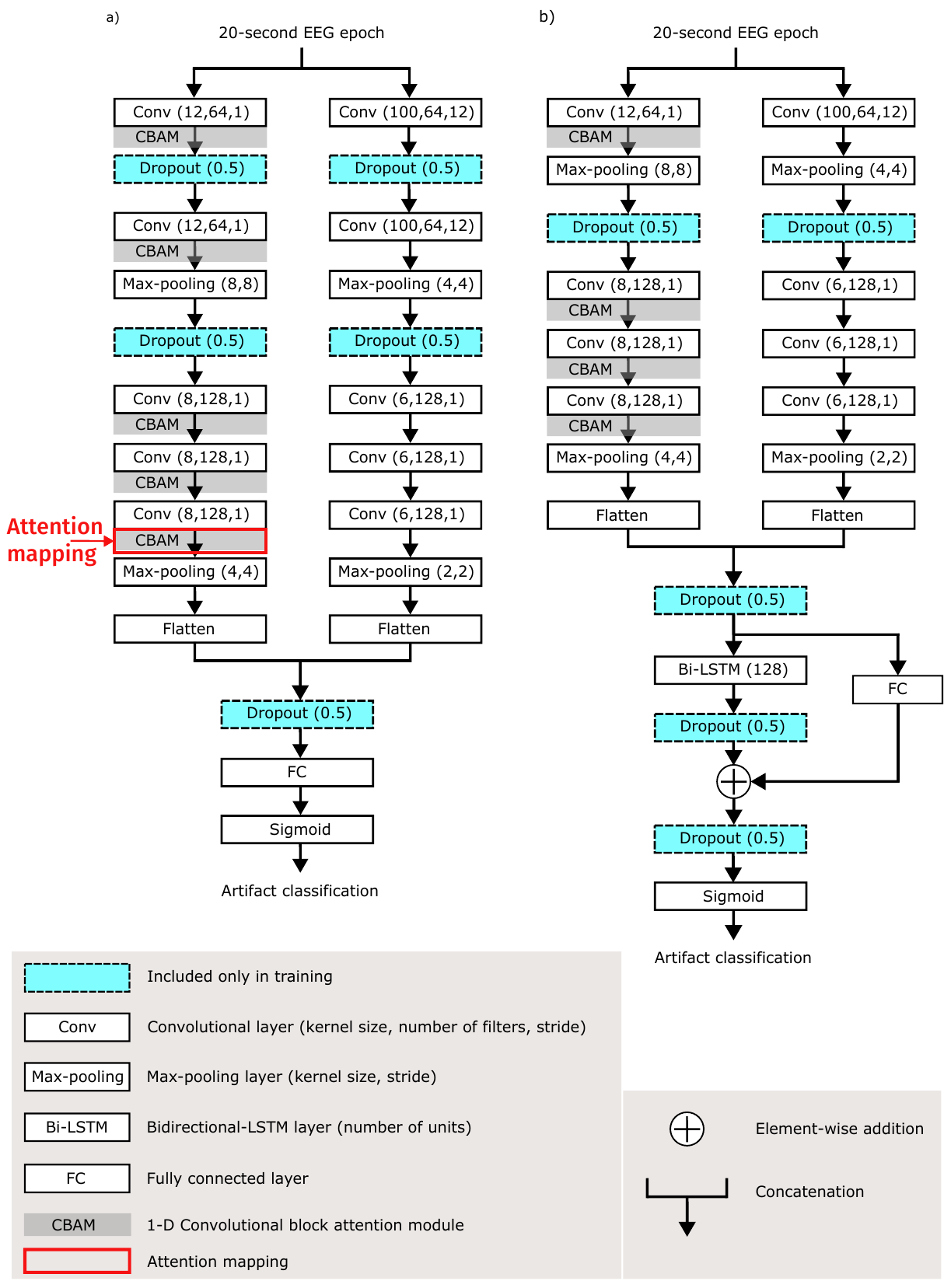}
    \caption{Deep learning models used for benchmarking: a) CNN and CNN-CBAM models, and b) CNN-LSTM and CNN-CBAM-LSTM models. The convolutional block attention module (CBAM) is applied to each convolutional layer on the temporal branch in the CNN-CBAM and CNN-CBAM-LSTM models. Dropout was applied only during training at a rate of 0.5 (blue). The attention mapping was performed after the CNN-CBAM model's last CBAM layer (red).}
    \label{fig:Figure_2}
\end{figure}

The CNN-LSTM model was built based on the same model for sleep stage classification with the same adaptations to the input and parameters described above \citep{dsn}. Two CNN branches learn spatial EEG features, followed by an LSTM that learns temporal information {\revisionc (Figure \ref{fig:Figure_2}.b)}. Similar to the model above, both branches include convolutional layers followed by batch normalization, dropout, and/or max pooling. The total number of convolutional layers in each branch is four. We replaced the two bidirectional LSTMs in the original model with a single bidirectional LSTM with 128 hidden units to reduce computational effort {\revisiond by 384,316 parameters (reduction by 27\%)}. To retain features from the CNN branches, a shortcut connection from the CNN outputs to the LSTM output was used. The LSTM layer is followed by a fully connected layer to complete the artifact classification.

\subsubsection{Convolutional block attention module on CNN and CNN-LSTM}

Convolutional layers fuse the spatial and channel-wise information by convolving with multichannel input or intermediate layers \citep{Hu2018}. {\revisionb To enhance the most critical and suppress less significant areas, we integrated a channel block attention module (CBAM) \citep{cbam}  into the above-described architectures. CBAM uses a sequential attention mechanism, applying channel and then spatial attentions one after another {\revisionc (Figure \ref{fig:Figure_3}).}

For this work, we adapted the well-known CBAM for time-series with 1D input. The 1D-CBAM was built sequentially in two steps: 1) channel attention and 2) spatial attention.
The channel attention of the CBAM captured the inter-channel relationship of the CNN feature maps. It squeezed the spatial information by applying both average-pooling (avg) and max-pooling (max) operations simultaneously along the spatial axis {\revisionc (Figure \ref{fig:Figure_3}, {\revisiond top})}. The generated descriptors $F^{c}_{avg}$ and $F^{c}_{max}$ were then fed into a shared layer with a  multi-layer perceptron with {\revisionb a reduction ratio of 8} {\revisiond obtained by fully connected (FC) layers and non-linear ReLU activation}. After the last shared FC layer, the two descriptors were summed up. A sigmoid activation function $ \sigma $ followed to obtain the channel-wise output $M_c(F)$, such as 
\begin{equation}
M_c(F) = \sigma(MLP(F^{c}_{avg}) + MLP(F^{c}_{max})).
\label{eq_channelMap}
\end{equation}
The spatial attention focused on finding where the information is located {\revisionc (Figure \ref{fig:Figure_3}, {\revisiond bottom})}. It was calculated as follows: 1) apply global average-pooling and max-pooling along the channel axis to generate two 1D feature maps $F^{s}_{avg}$ and $F^{s}_{max}$; 2) concatenate the two feature maps; 3) apply a convolution layer {\revisionb(kernel size = 7, number of filters = 1, stride = 1)} to the concatenated feature maps to generate the spatial attention map $M_s(F')$ such as 
\begin{equation}
M_s(F') = \sigma(f^{7}[F^{s}_{avg};F^{s}_{max}]),
\label{eq_spatialMap}
\end{equation}
where $ f^{7} $ was the 1D convolution with a 7$^th$ order filter. $M_s(F')$ was the spatial-wise output that encoded the regions {\revisionc along the time axis} in which the network emphasized informative or suppressed less informative characteristics. {\revisionc Woo et al. showed previously with an ablation study that the best performance is obtained when channel attention is arranged sequentially before the spatial attention module \citep{cbam}. Therefore, we adopted this architectural layout as well. }

\begin{figure}
    \centering
    \includegraphics[width=0.45\linewidth]{Figure_3_v2.pdf}
    \caption{{\revisionc The convolutional block attention module (CBAM) layer highlights the sequential connection of the channel- and spatial-wise attention in a convolutional neural network. The global average- and max-pooling are applied to the input features F to capture channel-wise dependencies. A shared fully connected (FC) layer forms a bottleneck, and a rectified linear unit (ReLU) function is used to introduce non-linearity. The channel-wise output  $M_c$ is computed by summing the output of the FC layer, then applying a sigmoid activation. Continuing in the sequence with the spatial attention, the average- and max-pooling are applied and concatenated to capture spatial dependencies. A convolution layer (Conv) is then applied, followed by a sigmoid function to generate the spatial-wise output  $M_s$.} }
    \label{fig:Figure_3}
\end{figure}

{\revisiond 
To translate the attention maps into temporal space, we conducted the activation attention mapping after the  CBAM in the last CNN layer of the temporal branch (Figure \ref{fig:Figure_2}a, red box). The activation attention mapping was proposed by Zagoruyko et al. to visualize the spatial attention map on CNN \citep{Zagoruyko2017}. We used the sum of absolute values raised to the power of \textit{p} as the activation-based mapping function, such as 
\begin{equation}
F_{sum}^{p}(A) = \sum_{i=1}^{C}|A_i|^p,
\label{eq:activation_att_map}
\end{equation}
where $A_i$ was the $i^{th}$ element of the feature map.  The value of \textit{p} places greater weight on the parts with the highest activations and can be adjusted for a specific task \citep{Zagoruyko2017}. {\revisionb It is the only parameter needed for fine-tuning.} The greater the value of \textit{p}, the more the model focuses on more significant parts, while suppressing less important ones. For this work, we have set \textit{p}=4 {\revisionb empirically to balance precision and recall. To ensure that at least 60\% of artifact windows were correctly classified, we set a minimum recall limit of 0.6. The selected value of p maximized precision while satisfying the specified recall requirements.} }      

We integrated the CBAM attention mechanism into the baseline CNN and CNN-LSTM models (CNN-CBAM and CNN-CBAM-LSTM, respectively). The attention mechanism was implemented after each CNN layer on the branch with the smaller kernel size (Figure \ref{fig:Figure_2}, {\revisiond grey boxes}), which is considered the temporal branch.

\subsection{{\revisiond Benchmark } algorithms}

{\revisionb To provide a meaningful baseline for comparison with our deep learning method, we selected three benchmark methods for artifact detection which are commonly used and available open-source: a spectral power threshold-based approach \citep{Huber2000, Leach2020}, because it is a widely used technique in a frequency domain{\revisiond;} a standard deviation threshold-based approach \citep{yasa}, because it is a typical amplitude-based artifact rejection algorithm{\revisiond;} and a heuristic-based 1D-CNN method \citep{Paissan2022}, which represents a recent machine learning approach that has demonstrated a strong performance compared to other CNN-based approaches for artifact detection.}

\subsubsection{Spectral power threshold-based detection algorithm}

First, the 50 Hz power-grid noise was removed from the raw EEG with a notch filter, followed by a band-pass between 0.5 and 40 Hz with a Butterworth band-pass filter. From this signal, we calculated the power spectral density using Welch's method for each 20-s epoch. A baseline threshold was calculated per night by averaging the power in the 0.75-4.5 Hz and 20-30 Hz bands from epochs that were manually scored as N1, N2, and N3 sleep stages according to the standard criteria \citep{Iber2007}. An artifact was detected when the power spectral density in an epoch exceeded this threshold. {\revisionb In contrast to our method, which aims to minimize preprocessing requirements, this approach includes several preprocessing steps like filtering, manual sleep scoring, and spectral power transformation.} It is essential to note that this approach is unsuitable for a real-time system as it requires manual sleep scoring and information on the full-night spectral power before it can be applied.

\subsubsection{Standard deviation threshold-based detection algorithm from YASA toolbox}

The standard deviation threshold-based algorithm is a standard approach in signal processing that defines outliers using standard deviations and a thresholding mechanism. This is a commonly used approach, available in various EEG frameworks, such as YASA Toolbox \citep{yasa}, MNE-Python \citep{mne}, and
EEGLAB \citep{eeglab}. To implement this algorithm, we used the Python Toolbox YASA v0.7.0. 

The YASA algorithm processes the whole recording at once, dividing it into small windows of predefined length. For each window, the standard deviation was first computed, and the resulting array was then log-transformed and z-scored. Windows with values exceeding the threshold were considered artifacts. We set the window length to 4 seconds to align with the same duration for artifact labels provided by experts. 
After predicting artifacts in each window, the windows were grouped into epochs. Each 20-second epoch consists of five consecutive 4-second windows. 
If at least one window within the epoch was detected as an artifact, the epoch was marked as an artifact. {\revisionb  While this approach is straightforward to implement, it depends heavily on the selected fixed threshold, making it less adaptable across different subjects and recording settings. As a result, it may require ongoing manual parameter tuning, which is not practical for real-time applications. Additionally, this algorithm lacks localization capability and therefore does not provide temporal information about the artifacts. }

\subsubsection{Heuristic-based 1D-CNN method}

Paissan et al. proposed a one-dimensional CNN architecture for detecting single-channel EEG artifacts and interpreting the frequency domain output feature maps  \citep{Paissan2022}. The model consists of a convolutional layer without downsampling, batch normalization, a rectified linear unit (ReLU) activation function, and global average pooling. The output is then passed through two fully connected layers with 8 and 3 hidden units, respectively. The softmax activation function is applied to the output of the last fully connected layer to generate probabilities. We used input vectors of a 20-second window and adapted the output to produce only two probabilities for artifact and non-artifact classes. 
In the original pipeline, the authors applied the Fourier transformation to the feature maps of the convolutional layer and computed their power spectral density to interpret the extracted features. This enabled the determination of the most critical frequencies used for classification. {\revisionb However, as our goal is to localize artifacts in the time domain, we use the model only for artifact classification.}  

\subsection{Experiments}

{\revisiond \subsubsection{Performance assessment} }

The performance was evaluated and compared using the area under the receiver operating characteristic (ROC) curve (AUC) metric. 
Additionally, we reported sensitivity ($se$) and specificity ($sp$)  for the best threshold on the ROC curve, such as
\begin{equation}
se = \frac{TP}{TP+FN}
\label{eq:sensitivity}
\end{equation}
and
\begin{equation}
sp = \frac{TN}{TN+FP}
\label{eq:specificity}
\end{equation}
where $TP$ denotes true positive, $FN$  false negative, $TN$  true negative, and $FP$  false positive. The best threshold was found by maximizing the geometric mean of $se$ and $sp$ {\revisionc (Equation \ref{eq:gmean})} {\revisiond on the validation set}, while changing the probability threshold for the predictions from 0 to 1 with an increment of 0.01.  
{\revisionc
\begin{equation}
gmean = \sqrt{se*sp}
\label{eq:gmean}
\end{equation}}

    
    For the threshold-based spectral power approach, only the $se$ and $sp$ were available for reporting.

 To assess the attention mechanism's ability to accurately predict artifact locations, we compared its predictions with manual labels for the 4-second windows. 
{\revisiond A prediction was considered correct if more than 50\% of its duration coincided with a manually labeled artifact window.}

We quantitatively evaluated the capacity of the attention mechanism to localize the artifact with the $se$ and $sp$ calculated over the manually labeled 4-s windows. {\revisiond We additionally reported the Jaccard index $J$ to capture the spatial overlap between the predicted and manually labeled 4-s windows, as it jointly penalizes both false positives and false negatives in a single measure. 
\begin{equation}
J = \frac{TP}{TP+FP+FN}
\label{eq:jaccard}
\end{equation}
}
We report these localization metrics for an ideal threshold, which we obtained by maximizing the geometric mean of the corresponding $se$ and $sp$ (see also \ref{sec:artifloc}). In addition to calculating performance metrics, we generated confusion matrices and conducted a qualitative analysis by plotting EEG segments alongside their corresponding attention maps. Using the ideal threshold, we visualized the artifacts identified by the attention mechanism in comparison to the ground-truth label locations.

{\revisiond \subsubsection{Optimization for artifact classification} }

We trained each of our deep learning models using the cross-entropy loss function and Adam optimizer {\revisionc with a learning rate of $5 \times 10^{-4}$} over 100 epochs. The batch size was set to 128. We applied {\revisionc kernel and activity regularizations, each with a coefficient of 0.001. Early stopping was implemented based on the total number of training epochs and empirical observations of the validation loss dynamics, with a patience of 20. {\revisiond Dropout at 0.5 was applied at 5 locations (2 per branch, 1 after concatenation for CNN-CBAM  (Figure \ref{fig:Figure_2}a), and 1 per branch, 1 after concatenation, Bi-LSTM layer, and element-wise addition, respectively for CNN-CBAM-LSTM (Figure \ref{fig:Figure_2}b).} Thus, we achieve training that reduces overfitting while preventing early termination due to validation fluctuations attributed to noise.} The model with the lowest loss 
was selected and tested with the testing set. 

The benchmark algorithms were optimized as follows: The spectral power threshold-based detection algorithm required no tuning as the threshold was based on the night's spectral content. We tuned the threshold for the YASA algorithm over the range 0.05 to 3 in increments of 0.01 to identify the optimal value by maximizing the geometric mean of $se$ and $sp$ for classifying 20-second epochs. We selected this threshold range because the results beyond this interval mirrored those at the edge. 
The one-dimensional CNN model for benchmarking was trained using a cross-entropy loss function and optimized with a double strategy. Two Adam optimizers were used during training: first to learn the weights of the convolutional part and, second, to fine-tune the weights of the two fully connected layers. {\revisionc The learning rates were specified as 0.01 and 0.001, respectively. Gradients were computed and applied separately to the convolutional and fully connected parts at every training step, enabling layer-specific optimization. We also applied L2 regularization with a value of 0.01.} The training was interrupted if the model did not improve within 20 consecutive learning epochs.  
\\
{\revisiond \subsubsection{Characterization of artifact localization}  \label{sec:artifloc}
Following the initial epoch-level classification, we isolated all 20-second segments predicted by the model to contain artifacts, which naturally pooled both TP and FP classifications for downstream analysis. Because no separate holdout validation set or independent test set was designated for this secondary localization task, the threshold search was performed directly on the combined pool of predicted segments. For these selected segments, we min-max scaled the activation-based attention maps to a standard [0,1] range. To prevent convolutional edge effects from introducing boundary distortions, we explicitly excluded the first and last 0.7 seconds of each segment from the localization evaluation. The resulting activation maps resolved the relative intensity of signal anomalies continuously along the time axis. We then applied a scalar threshold to these maps to obtain a discrete, binary indication of the artifact boundaries. Time points at which the attention map exceeded this threshold were designated as the exact artifact locations isolated by the attention mechanism. Crucially, because this threshold was determined directly on the same predicted segments under evaluation to maximize the geometric mean of localization sensitivity and specificity, these metrics represent an upper bound on performance for this specific dataset and allow fair comparison with other methods, such as YASA. However, the localization performance remains operationally dependent on the true-to-false-positive distribution produced by the primary epoch classifier, and this approach does not verify its generalizability across independent datasets.
}

\section{Results}
\subsection{Artifact classification {\revisiond performance}}
The CNN-CBAM model outperformed the other three deep learning architectures, achieving the highest classification metrics with an AUC of 0.88, a $se$ of 0.81, and a $sp$ of 0.86 (Table \ref{table:2}). {\revisiond Notably, this performance gain was achieved with a parameter count of 1,080,456, which remained comparable to the CNN model (1,065,474) and substantially lower than both LSTM-based models (1,415,362 and 1,425,286), suggesting that the CBAM attention mechanism improves classification without a significant increase in model complexity.} The 1D-CNN achieved a competitive AUC of 0.85 with only 426 parameters. Visual analysis of the aggregated ROC curves (Figure \ref{fig:Figure_5}) confirms that models incorporating CBAM layers consistently achieved higher performance than those without.
Furthermore, the CNN-CBAM model demonstrated superior classification capabilities when compared to established open-source algorithms, yielding higher ROC AUC, $se$, and $sp$ across the board.
Consequently, due to its best performance in the classification phase, we selected the CNN-CBAM model as the primary architecture for the subsequent artifact localization tasks.


\begin{table}[!htb]
\caption{\small The models' performances with the area under the curve (AUC), sensitivity ($se$), and specificity ($sp$) for the four deep learning models {\revisionb  (ablation study) } and the open-source benchmark algorithms. The bold number indicates the highest performance for each metric.}
\centering
\setlength{\tabcolsep}{+4pt}
\setlength\extrarowheight{+3pt}
\begin{tabular}{p{150pt}|C{50pt}|C{25pt}C{50pt}C{25pt}}
\hline
\small \textbf{Model} 		 & Parameter count 				& \small $AUC$ 			 & \small $se$          & \small $sp$ 	\\ [4pt]
\hline
\small CNN					 &	1,065,474	&0.73			 &0.66 			  &0.68 \\
\small CNN-LSTM 		    &	1,415,362	&0.77			 &0.66 			  &0.80 \\
\small CNN-CBAM	    		 &	1,080,456 &\textbf{0.88} & \textbf{0.81}  & \textbf{0.86} \\
\small CNN-CBAM-LSTM		 &	1,425,286	&0.84			 &0.78 			  &0.82 \\	\hline
\small \textbf{Benchmarks}  & & &\\
\small Spectral power approach$^{*}$     & -  & -              & 0.35          & 0.79 \\
\small Standard deviation approach {\revisiond(YASA)}            & -           &0.72            & 0.64          & 0.69 \\
\small 1D-CNN              & 426          &0.85            & 0.76          & 0.81 \\
\hline 
\multicolumn{5}{p{280pt}}{\small $^{*}$ unable to classify artifacts in REM sleep and Wake stages }\\
\multicolumn{5}{p{280pt}}{}\\
\end{tabular}
\label{table:2}
\end{table}

\begin{figure}[!h]
    \centering
    \includegraphics[width=0.55\linewidth]{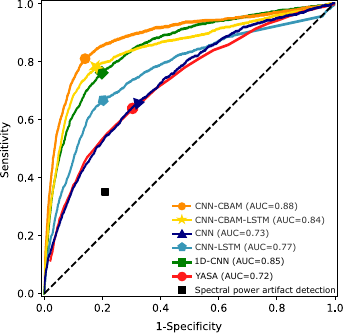}
    \caption{Aggregated receiver operating characteristic (ROC) and the reported area under the curve (AUC) for the four deep learning models in the ablation study (CNN-*) and the three open-source benchmark algorithms. The dots on each ROC curve denote the best threshold with a balanced trade-off between $se$ and $sp$ as reported in Table \ref{table:2}. The dashed line indicates the performance of a uniform random guess classifier.}
    \label{fig:Figure_5}
\end{figure}

\begin{figure}
    \centering
    \includegraphics[width=1\linewidth]{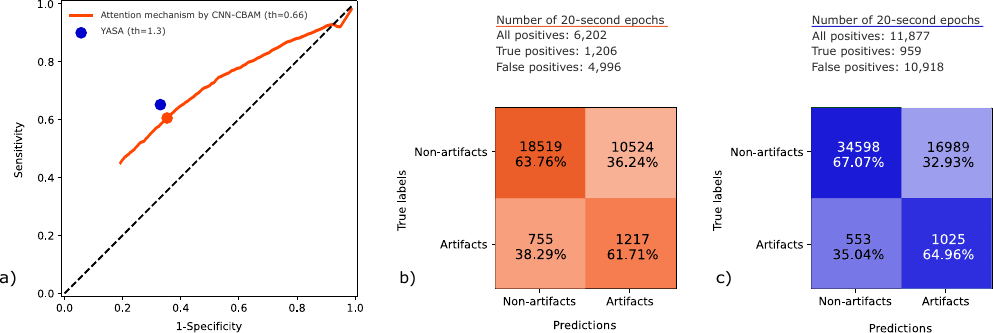}
    \caption{Comparison of the attention mechanism of our CNN-CBAM model and an open-source threshold-based standard deviation approach using YASA Toolbox (YASA) for artifact localization: a) Aggregated receiver operating characteristics (ROC) for the attention mechanism of CNN-CBAM and YASA (blue point). The dashed line indicates the performance of a uniform random guess classifier. \textit{th} denotes threshold. Confusion matrices illustrate the quantitative performance of each method, showing the percentage of correctly predicted artifacts and non-artifacts, along with the number of 20-second epochs (all positives) used in each case for b) the CNN-CBAM model and c) YASA.}
    \label{fig:Figure_6}
\end{figure}

\begin{figure}[!h]
    \centering
    \includegraphics[width=0.98\linewidth]{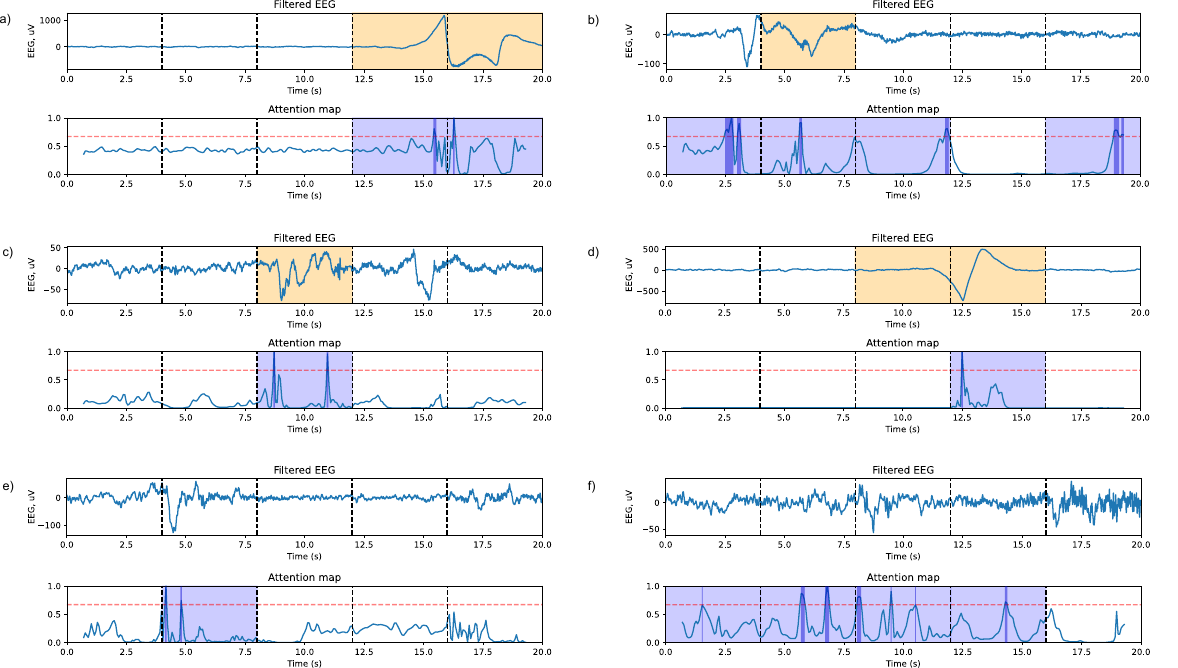}
    \caption{Visualization of artifacts and attention maps for 6 exemplary 20-second epochs classified as containing artifacts (a-f). Top: Filtered EEG signal, presented to the experts, with manually scored artifacts in 4-second windows (orange shaded area). Bottom: Normalized attention maps divided into five 4-second  windows. The attention map, which exceeded the threshold of 0.66 (dashed red line), classified windows with artifacts (blue shaded area) and identified the time points of artifact occurrence (dark blue shaded area).}
    \label{fig:Figure_7}
\end{figure}

\subsection{Artifact localization {\revisiond performance}}
The attention mechanism of the CNN-CBAM achieved a $se$ of  0.61 and a $sp$ of 0.63, with an optimal threshold at 0.66 (Figure \ref{fig:Figure_6}, a). For comparison, the standard deviation approach within the YASA Toolbox yielded a slightly higher $se$ of 0.64 and $sp$ of 0.67 at an optimal threshold of 1.3. {\revisiond However, $J$ was 9.74\% for CNN-CBAM and 5.52\% for YASA, indicating that both methods produced a substantial number of false positives at the 4-sec window level, with CNN-CBAM achieving nearly twice the overlap accuracy of YASA. This discrepancy between $sp$ and $J$ is further explained by the confusion matrices (Figure \ref{fig:Figure_6}, b-c), which revealed that YASA's performance was more significantly compromised by the initial misclassification of 20-second EEG epochs, which led to a higher number of false positives in the localization step.}
To qualitatively validate these findings, we visualized the CNN-CBAM attention maps. These visualizations demonstrate how the attention mechanism aligns with manual labels to precisely identify the temporal occurrence of artifacts (Figure \ref{fig:Figure_7}).

\section{Discussion}
We proposed four end-to-end deep learning models to classify and visualize artifacts in 20-second EEG epochs recorded with a wearable device in an uncontrolled sleep setting. The four models were based on a CNN and a CNN-LSTM architecture, evaluated both with and without CBAM layers. We demonstrated that CBAM-based models achieve higher performance than those without an integrated attention mechanism. This highlights the importance of guiding the network's learning toward the most informative components of the raw input signal. All deep learning models outperformed traditional feature-based signal processing algorithms that rely on spectral power or standard deviation thresholds, achieving performance comparable to or superior to a heuristic-based 1D-CNN approach. 

{\revisiond A major observation from our ablation study is that the smaller CNN-CBAM model outperformed the CNN-CBAM-LSTM model, which incorporates recurrent layers. While intuition suggests that processing temporal features via recurrent layers should yield better performance on time-series data, this was not the case in our experiments. An explanation for this outcome could be that the CBAM attention mechanism already successfully focuses on the most relevant temporal and spatial features using convolutions. Adding a  Bi-LSTM layer on top may introduce parameter redundancy and unnecessary architectural complexity, ultimately reducing performance. Indeed, further parameter analysis revealed a clear trade-off between model complexity and classification performance. Despite using only 426 parameters, the benchmark 1D-CNN achieved a competitive AUC of 0.85, compared to 0.88 for the CNN-CBAM model, which required little more than 1 million parameters. This suggests that simpler architectures may be sufficient if the objective is strictly artifact classification, which could be an advantage for direct inference on mobile devices with low computational power. However, larger models equipped with attention mechanisms provide deeper insights. In our framework, the CNN-CBAM model uniquely enabled artifact localization through its attention mechanism, allowing for granular identification of noise within the 20-second epochs. } 

The processing and interpretation of an EEG contaminated with artifacts are highly challenging tasks. While traditional signal processing approaches must be manually tailored to specific artifact types, time scales, and signal sources, deep learning can automate this process. In this work, we demonstrated that CNNs are powerful tools for classifying artifacts in physiological signals. Featuring a two-branch CNN, our models learned both temporal and frequency information directly from raw EEG data without requiring sophisticated preprocessing or manual feature engineering  \citep{Bahador2020}. 
 

Crucially, the CNN-CBAM model enabled the visualization of artifacts from a single EEG channel. The integrated attention module evaluated both channel-wise and spatial-wise regions of the EEG, providing clear insight into which signal sections contributed to a positive artifact classification. The qualitative evaluation of the CNN-CBAM attention map showed that the high-amplitude attention regions closely matched the reference labels used in manual sleep scoring. However, this alignment was not perfect, yielding a localization $se$ of 0.61. Mismatches were frequently located at adjacent windows. A likely explanation for this is that manual labeling is tedious and highly prone to inconsistencies. Labels can be strongly biased toward individual human scorers, which introduces substantial inter-rater variability. Furthermore, labeling artifacts in sleep EEG can be biased based on the clinical or research objective \citep{Malafeev2019}, meaning that minor artifacts less relevant to the immediate task might go unnoticed by human experts and impact sleep staging.


{\revisiond Despite these human limitations, the CNN-CBAM model outperformed the standard YASA toolbox approach in artifact localization, as reflected by a nearly doubled Jaccard index  ($J$: 9.74\% vs. 5.52\%). This suggests that the attention-based localization strategy produces fewer incidental or false artifact detections. While YASA achieved slightly higher raw $se$ and $sp$ at the window level, its localization performance was substantially degraded by false positives propagating directly from epoch-level misclassifications. This highlights the clear advantage of an end-to-end learned localization strategy. }
We visually inspected the locations of the artifacts identified by the attention mechanism to qualitatively validate these findings. For instance, while the manual expert label completely missed an artifact that appeared around 4.5 seconds in Figure \ref{fig:Figure_7}.e, the attention map successfully caught the noise by displaying a high activation probability at that exact temporal location. Implementing an automated artifact detection algorithm with an attention mechanism could strongly facilitate human scoring workflows by providing reliable, objective, pre-annotated artifact labels. This integration of attention mechanisms generates significant interest in semi-automated decision-making, driving greater transparency and trust in machine learning models that require human interpretation. Ultimately, the attention-assisted visualization of artifacts in quasi-real-time can boost the clinical adoption of medical wearables and IoMT applications.



{\revisionb Artifact detection is inherently difficult because the number of epochs contaminated with noise is usually much smaller than the number of clean epochs, creating a severe class imbalance. To mitigate this issue within our wearable EEG dataset, we generated synthetic data using SMOTE.
{\revisiond Although SMOTE has been previously extended to time-series applications \citep{yadav2025, Wang2024}, it was not originally designed for this data type. Because it generates synthetic samples by linearly interpolating between existing instances, the resulting segments may contain unrealistic patterns that do not fully reflect the morphology of true physiological artifacts. To assess the plausibility of the generated samples, we conducted a qualitative comparison between original and synthetic artifact segments using PSD analysis. This analysis confirmed that synthetic samples preserved the general spectral and morphological characteristics of real artifact epochs (Appendix~\ref{appendix:A}). 
Nevertheless, this validation remains qualitative, and establishing a rigorous assessment of synthetic sample quality remains an open challenge. Additionally, including synthetic data in the validation set stabilizes threshold selection during training but risks biasing parameter tuning toward the minority class, potentially distorting true performance metrics. In future designs, it would be highly desirable to integrate dedicated cost-sensitive or architectural mechanisms to address class imbalance directly within the deep learning network.}

The defined rules for correct artifact localization provided a reliable detection paradigm in general. {\revision However, the ideal attention threshold used for reporting was obtained on the same segments predicted as artifact-contaminated, which may introduce a bias toward this specific dataset. Independent validation on an external dataset is required to confirm the generalizability of the selected threshold fully. } Furthermore, mathematical edge cases exist: when an artifact spans two adjacent windows, or when a window contains partial artifact activity that does not exceed half of the window size, it will fail to meet the 50\% overlap threshold. In those cases, the segment will be classified as artifact-free despite containing low-level noise.


While the attention mechanism allows us to identify the exact time points at which artifacts occur, estimating the absolute duration depends heavily on the temporal resolution. With our specific downsampling and architectural design, we were limited to an estimated shortest duration of 60 ms. If a shorter transient artifact occurred, the mechanism would still flag it, but the estimated duration would be recorded as 0 seconds. Due to the lack of exact start and end timestamps in the expert ground truth, we could not quantitatively evaluate the mechanism's precision in boundary localization.


Finally, labeling for the available dataset was strictly limited to the operational needs of sleep stage scoring. Consequently, less relevant artifacts, such as those occurring during wake stages, low-frequency sweating artifacts, or cardiogenic activity, were intentionally left unlabeled. Furthermore, the findings of this study are based entirely on a dataset from a highly specific study population of healthy older adults, which limits direct generalizability to broader clinical cohorts or younger populations. Future work must address the quality and reliability of reference data by sourcing it from a more diverse population.

Unlike traditional spectral power threshold-based approaches, our proposed deep learning models do not require prior information about individual subject baselines or manual sleep stage scoring. Therefore, they are highly suitable for deployment in quasi-real-time monitoring systems. By dynamically adjusting the attention threshold, human raters can intuitively tune the detection sensitivity to match their specific analytical preferences.

Expanding this framework to utilize fully categorized artifact labels alongside a multi-class neural network output could significantly increase specificity. This would enable highly specialized artifact classifiers, such as those dedicated to isolating cardiogenic artifacts \citep{Chiu2021}. Such specialized classifiers would allow additional clean physiological information to be extracted, expanding applications beyond sleep staging into daytime EEG recordings, brain-computer interfaces (BCIs), and complex multi-channel EEG setups where real-time feedback is critical. However, manual labeling of multi-channel artifacts is exceptionally costly and can rapidly become unfeasible. To address this data bottleneck, more research on reliable training strategies with sparse labels is required. One promising approach our group has introduced for this task is distantly supervised multitask learning networks, which successfully reduce the volume of required manual labels by incorporating multiple related auxiliary tasks during training \citep{Schwab2018b}.


There are distinct architectural pathways for implementing these automated models in real-world scenarios. Applying such models at the edge could enable real-time user interaction to improve recording quality, for instance, by triggering user "nudges" to adjust a loose electrode band. Alternatively, a cloud-based implementation could drastically enhance large-scale wearable data collection by enabling remote clinical supervision to ensure data integrity and track patient adherence. Early warnings could be automatically triggered when cumulative artifact levels exceed predefined target thresholds for specific patients. Furthermore, cloud infrastructure would enable parallel computation of multiple auxiliary algorithms, such as the YASA toolbox, alongside our deep learning framework to provide comprehensive secondary analytics. Clinical staff could then visually review this multi-modal quality profile and promptly implement targeted interventions to maximize long-term patient adherence and data trust.

\section{Conclusion}
{\revisiond We proposed a deep learning framework for detecting artifacts in EEG data collected by wearable sleep devices. An ablation study revealed that the CNN-CBAM model outperformed the alternative CNN, CNN-LSTM, and CNN-CBAM-LSTM architectures, as well as the established open-source algorithms based on spectral signal processing. This research demonstrates a novel and effective approach for quasi-real-time artifact detection in raw, single-channel EEG signals. Crucially, the system eliminates the labour-intensive process of manual feature engineering. We further demonstrate that integrating attention maps with a single, tunable parameter provides a mechanism for artifact localization, enabling the rapid visualization of artifacts. This streamlined approach ensures both technical simplicity and high practical utility for real-world applications.}

\section*{Conflict of Interest Statement}

The authors declare that the research was conducted without any commercial or financial relationships that could be construed as a potential conflict of interest.

\section*{Acknowledgments}
We thank Luzius Brogli for the meaningful discussions on developing the classification models and the members of Leitwert AG for the many discussions about potential implementations. We thank the SleepLoop consortia members who contributed to creating the data set: Caroline Lustenberger managed the clinical trial and data collection. Reto Huber, Stephanie Huwiler, Esther Werth, and Caroline Lustenberger contributed to the labeling of artifacts and sleep stages. Renato Büchi, Nino Demarmels, Gary Hoppeler, Jérôme Kurz, and Eva Silberschmidt supported the recruitment and data collection. We thank all participants for volunteering in the study and sharing their data.

\section*{Author Contributions}
All authors conceived and designed the project and revised and approved the final version of the manuscript. WK secured funding. KS, JZ, and WK drafted the manuscript.  JZ designed the machine learning models, conducted the experiments, and analyzed the data. KS reproduced and conducted experiments, validated the findings, performed a benchmarking study with open-source approaches, and analyzed the data. 
MLF designed the device for data collection. 

\section*{Funding}
This work was partially funded by InnoSuisse (Grant No. 35484.1, 2018-2020), in collaboration with Leitwert AG, and by financial support programs for female researchers, Office for Gender Equality, Ulm University. Data collection and management were conducted as part of SleepLoop, a Flagship of Hochschulmedizin Zürich. 

\section*{Declaration of generative AI and AI-assisted technologies in the writing process}
During the preparation of this work, the authors used Grammarly v14.1208.0 to check grammar, punctuation, and spelling. All names were removed or changed while passing the sentences to Grammarly. After using this tool, the authors reviewed and edited the content as needed and take full responsibility for the published article.


\bibliographystyle{ieeetr}
\bibliography{artfactEEG_re_v3}

@article{dsn,
author = {Supratak, Akara and Dong, Hao and Wu, Chao and Guo, Yike},
doi = {10.1109/TNSRE.2017.2721116},
issn = {15344320},
journal = {IEEE Transactions on Neural Systems and Rehabilitation Engineering},
number = {11},
pages = {1998--2008},
title = {{DeepSleepNet: A model for automatic sleep stage scoring based on raw single-channel EEG}},
volume = {25},
year = {2017}
}

@INPROCEEDINGS{sadiya2021,
  author={Sadiya, Sari and Alhanai, Tuka and Ghassemi, Mohammad M},
  booktitle={2021 10th International IEEE/EMBS Conference on Neural Engineering (NER)}, 
  title={Artifact Detection and Correction in EEG data: A Review}, 
  year={2021},
  volume={},
  number={},
  pages={495-498},
  doi={10.1109/NER49283.2021.9441341}}

@article{islam2016,
title = {Methods for artifact detection and removal from scalp EEG: A review},
journal = {Neurophysiologie Clinique/Clinical Neurophysiology},
volume = {46},
number = {4},
pages = {287-305},
year = {2016},
issn = {0987-7053},
doi = {https://doi.org/10.1016/j.neucli.2016.07.002},
url = {https://www.sciencedirect.com/science/article/pii/S098770531630199X},
author = {Md Kafiul Islam and Amir Rastegarnia and Zhi Yang},

}

@ARTICLE{mannan2018,
  author={Mannan, Malik Muhammad Naeem and Kamran, Muhammad Ahmad and Jeong, Myung Yung},
  journal={IEEE Access}, 
  title={Identification and Removal of Physiological Artifacts From Electroencephalogram Signals: A Review}, 
  year={2018},
  volume={6},
  number={},
  pages={30630-30652},
  doi={10.1109/ACCESS.2018.2842082}}

@InProceedings{cbam,
author="Woo, Sanghyun
and Park, Jongchan
and Lee, Joon-Young
and Kweon, In So",
editor="Ferrari, Vittorio
and Hebert, Martial
and Sminchisescu, Cristian
and Weiss, Yair",
title="CBAM: Convolutional Block Attention Module",
booktitle="Computer Vision -- ECCV 2018",
year="2018",
publisher="Springer International Publishing",
address="Cham",
pages="3--19",
isbn="978-3-030-01234-2"
}

@article{Lustenberger2022,
  title={Auditory deep sleep stimulation in older adults at home: a randomized crossover trial},
  author={Lustenberger, C and Ferster, ML and Huwiler, S and Brogli, L and Werth, E and Huber, R and Karlen, W},
  journal={Commun Med},
  volume={2},
  pages={30},
  year={2022},
  publisher={Nature Publishing Group},
  doi={10.1038/s43856-022-00096-6},
}

@article{Ferster2019,
author = {Ferster, Maria Laura and Lustenberger, Caroline and Karlen, Walter},
doi = {10.1109/LSENS.2019.2914425},
journal = {IEEE Sensors Letters},
pages = {1--4},
title = {{Configurable Mobile System for Autonomous High-Quality Sleep Monitoring and Closed-Loop Acoustic Stimulation}},
volume={3},
number={5},
year = {2019}
}

@article{Raduntz2018,
author = {Radüntz, Thea},
doi = {10.3389/fphys.2018.00098},
journal = {Frontiers in Physiology},
pages = {1--12},
title = {{Signal quality evaluation of emerging EEG devices}},
volume={9},
year = {2018}
}

@article{smote2002,
author = {Chawla, NV and Bowyer, KW and Hall, LO and Kegelmeyer, WP},
doi = {10.1613/jair.953},
journal = {Journal of Artificial Intelligence Research},
pages = {321--357},
title = {{SMOTE: Synthetic Minority Over-sampling Technique}},
volume={16},
year = {2002}
}

@article{Teplan2002,
author = {Teplan, Michal},
doi = {10.1021/pr070350l},
journal = {Measurement Science Review},
pages = {1--11},
title = {{Fundamentals of EEG measurement}},
volume={2},
year = {2002}
}

@article{Burger2015,
author = {Burger, Christiaan and Van Den Heever, David Jacobus},
doi = {10.1016/j.bspc.2014.09.009},
journal = {Biomedical Signal Processing and Control},
pages = {67--79},
title = {{Removal of EOG artefacts by combining wavelet neural network and independent component analysis}},
volume={15},
year = {2015}
}

@article{Rozario2015,
author = {D’Rozario, Angela L. and Dungan, George C. and Banks, Siobhan and Liu, Peter Y. and Wong, Keith K.H. and Killick, Roo and Grunstein, Ronald R. and Kim, Jong Won},
doi = {10.1007/s11325-014-1056-z},
journal = {Sleep and Breathing},
pages = {607--615},
title = {{An automated algorithm to identify and reject artefacts for quantitative EEG analysis during sleep in patients with sleep-disordered breathing}},
volume={19},
number={2},
year = {2015}
}

@article{Regan2013,
author = {O'Regan, Simon and Faul, Stephen and Marnane, William},
doi = {10.1016/j.medengphy.2012.08.017},
journal = {Medical Engineering and Physics},
pages = {867--874},
title = {{Automatic detection of EEG artefacts arising from head movements using EEG and gyroscope signals}},
volume={35},
number={7},
year = {2013}
}

@article{Wang2017,
author={Wang, F. and Jiang, M. and Qian, C. and Yang, S. and Li, C. and Zhang, H. and Wang, X. and Tang, X.},
doi = {10.1109/CVPR.2017.683},
journal = {IEEE Conference on Computer Vision and Pattern Recognition},
title = {{Residual attention network for image classification}},
pages={6450--6458},
year = {2017}
}

@article{Hu2018,
author={Hu, J. and Shen, L. and Sun, G.},
doi = {10.1109/CVPR.2018.00745},
journal = {IEEE/CVF Conference on Computer Vision and Pattern Recognition},
title = {{Squeeze-and-excitation networks}},
pages={7132--7141},
year = {2018}
}

@article{Casson2019,
author = {Casson, Alexander J.},
doi = {10.1007/s13534-018-00093-6},
journal = {Biomedical Engineering Letters},
pages = {53--71},
title = {{Wearable EEG and beyond}},
volume={9},
number={1},
year = {2019}
}

@article{Lebedev2017,
author = {Lebedev, MA and Nicolelis, MA.},
doi = {10.1152/physrev.00027.2016},
journal = {Physiol. Rev.},
pages = {767--837},
title = {{Brain-Machine Interfaces: From Basic Science to Neuroprostheses and Neurorehabilitation}},
volume={97},
number={2},
year = {2017}
}

@article{Zeng2018,
author = {Zeng, Hong and Yang, Chen and Dai, Guojun and Qin, Feiwei and Zhang, Jianhai and Kong, Wanzeng},
doi = {10.1007/s11571-018-9496-y},
journal = {Cognitive Neurodynamics},
pages = {597--606},
title = {{EEG classification of driver mental states by deep learning Hong}},
volume={12},
number={6},
year = {2018}
}

@article{Zarjam2015,
author = {Zarjam, Pega and Epps, Julien and Lovell, Nigel H.},
doi = {10.1109/TAMD.2015.2441960},
journal = {IEEE Transactions on Autonomous Mental Development},
pages = {301--310},
title = {{Beyond Subjective Self-Rating: EEG Signal Classification of Cognitive Workload}},
volume={7},
number={4},
year = {2015}
}

@article{Patanaik2018,
author = {Patanaik, Amiya and Ong, Ju Lynn and Gooley, Joshua J. and Ancoli-Israel, Sonia and Chee, Michael W.L.},
doi = {10.1093/sleep/zsy041},
journal = {Sleep},
pages = {1--11},
title = {{An end-to-end framework for real-time automatic sleep stage classification}},
volume={41},
number={5},
year = {2018}
}

@article{Tsinalis2016,
author = {Tsinalis, Orestis and Matthews, Paul M. and Guo, Yike},
doi = {10.1007/s10439-015-1444-y},
journal = {Annals of Biomedical Engineering},
pages = {1587--1597},
title = {{Automatic Sleep Stage Scoring Using Time-Frequency Analysis and Stacked Sparse Autoencoders}},
volume={44},
number={5},
year = {2016}
}

@article{Nathan2016,
author = {Nathan, Kevin and Contreras-Vidal, Jose L.},
doi = {10.3389/fnhum.2015.00708},
journal = {Frontiers in Human Neuroscience},
pages = {1--12},
title = {{Negligible motion artifacts in scalp electroencephalography (EEG) during treadmill walking}},
volume={9},
year = {2016}
}

@article{Kilicarslan2016,
  author    = {Atilla Kilicarslan and Robert G. Grossman and Jose Luis Contreras-Vidal},
  title     = {A Robust Adaptive Denoising Framework for Real-Time Artifact Removal in Scalp EEG Measurements},
  journal   = {Journal of Neural Engineering},
  year      = {2016},
  volume    = {13},
  number    = {2},
  pages     = {026013},
  doi       = {10.1088/1741-2560/13/2/026013},
}

@article{Craik2019,
doi = {10.1088/1741-2552/ab0ab5},
url = {https://doi.org/10.1088/1741-2552/ab0ab5},
year = {2019},
publisher = {IOP Publishing},
volume = {16},
number = {3},
pages = {031001},
author = {Craik, Alexander and He, Yongtian and Contreras-Vidal, Jose L},
title = {Deep learning for electroencephalogram (EEG) classification tasks: a review},
journal = {Journal of Neural Engineering},
}

@inproceedings{
Zagoruyko2017,
title={Paying More Attention to Attention: Improving the Performance of Convolutional Neural Networks via Attention Transfer},
author={Sergey Zagoruyko and Nikos Komodakis},
booktitle={5th International Conference on Learning Representations, ICLR 2017},
year={2017},

}

@article{Leach2020,
  author    = {Sven Leach and Ku-Young Chung and Laura Tüshaus and Reto Huber and Walter Karlen},
  title     = {A Protocol for Comparing Dry and Wet {{EEG}} Electrodes During Sleep},
  journal   = {Frontiers in Neuroscience},
  year      = {2020},
  volume    = {14},
  pages     = {586},
  doi       = {10.3389/fnins.2020.00586},
  publisher = {Frontiers Media SA},
}

@article{Li2015,
author = {Li, Gang and Lee, Boon Leng and Chung, Wan Young},
doi = {10.1109/JSEN.2015.2473679},
journal = {IEEE Sensors Journal},
title = {{Smartwatch-Based Wearable EEG System for Driver Drowsiness Detection}},
volume={15},
number={12},
pages={7169-7180},
year = {2015}
}

@article{Sterr2018,
author = {Sterr, Annette and Ebajemito, James K. and Mikkelsen, Kaare B. and Bonmati-Carrion, Maria A. and Santhi, Nayantara and della Monica, Ciro and Grainger, Lucinda and Atzori, Giuseppe and Revell, Victoria and Debener, Stefan and Dijk, Derk Jan and DeVos, Maarten},
doi = {10.3389/fnhum.2018.00452},
journal = {Frontiers in Human Neuroscience},
title = {{Sleep EEG derived from behind-the-ear electrodes (cEEGrid) compared to standard polysomnography: A proof of concept study}},
volume={12},
number={November},
pages={1-9},
year = {2018}
}

@INPROCEEDINGS{Bahador2020,

  author={Bahador, Nooshin and Erikson, Kristo and Laurila, Jouko and Koskenkari, Juha and Ala-Kokko, Tero and Kortelainen, Jukka},

  booktitle={2020 42nd Annual International Conference of the IEEE Engineering in Medicine \& Biology Society (EMBC)}, 

  title={Automatic detection of artifacts in EEG by combining deep learning and histogram contour processing}, 

  year={2020},

  volume={},

  number={},

  pages={138-141},

  doi={10.1109/EMBC44109.2020.9175711}}

@article{Huber2000,
author = {Huber, R. and Graf, T. and Cote, K. A. and Wittmann, L. and Gallmann, E.
 and Matter, D. and Schuderer, J. and Kuster, N. and Borbely, A. A. and Achermann, P.},
doi = {10.1097/00001756-200010200-00012},
journal = {NeuroReport},
title = {{Exposure to pulsed high-frequency electromagnetic field during waking affects human sleep EEG}},
pages={3321--3325},
volume = {11},
number = {15},
year = {2000}
}

@book{Iber2007,
  title={The AASM Manual for the Scoring of Sleep and Associated Events: Rules, Terminology and Technical Specifications},
  author={Iber, C. and Ancoli-Israel, S. and Chesson, A. and Quan, S. F.},
  year={2007},
  publisher={American Academy of Sleep Medicine},
  address={Westchester, IL}
}

@InProceedings{Schwab2018b,
  title = 	 {Not to Cry Wolf: Distantly Supervised Multitask Learning in Critical Care},
  author =       {Schwab, Patrick and Keller, Emanuela and Muroi, Carl and Mack, David J. and Str{\"a}ssle, Christian and Karlen, Walter},
  booktitle = 	 {Proceedings of the 35th International Conference on Machine Learning},
  pages = 	 {4518--4527},
  year = 	 {2018},
  volume = 	 {80},
}

@article{Chiu2021,
title = {Get rid of the beat in mobile EEG applications: A framework towards automated cardiogenic artifact detection and removal in single-channel EEG},
journal = {Biomedical Signal Processing and Control},
volume = {72},
pages = {103220},
year = {2022},
issn = {1746-8094},
doi = {https://doi.org/10.1016/j.bspc.2021.103220},
author = {Neng-Tai Chiu and Stephanie Huwiler and M. Laura Ferster and Walter Karlen and Hau-Tieng Wu and Caroline Lustenberger},
}

@INPROCEEDINGS{peh2022,

  author={Peh, Wei Yan and Yao, Yuanyuan and Dauwels, Justin},

  booktitle={2022 44th Annual International Conference of the IEEE Engineering in Medicine \& Biology Society (EMBC)}, 

  title={Transformer Convolutional Neural Networks for Automated Artifact Detection in Scalp EEG}, 

  year={2022},

  volume={},

  number={},

  pages={3599-3602},

  doi={10.1109/EMBC48229.2022.9871916}}

@article{saba2021,
  title={Unsupervised EEG Artifact Detection and Correction},
  author={Saba-Sadiya, Sari and Chantland, Eric and Alhanai, Tuka and Liu, Taosheng and Ghassemi, Mohammad M.},
  journal={Frontiers in Digital Health},
  volume={2},
  pages={608920},
  year={2021},
  publisher={Frontiers Media SA},
  doi={10.3389/fdgth.2020.608920}
}

@INPROCEEDINGS{Paissan2022,
  author={Paissan, Francesco and Kumaravel, Velu Prabhakar and Farella, Elisabetta},
  booktitle={2022 IEEE Sensors Applications Symposium (SAS)}, 
  title={Interpretable CNN for Single-Channel Artifacts Detection in Raw EEG Signals}, 
  year={2022},
  volume={},
  number={},
  pages={1-6},
  doi={10.1109/SAS54819.2022.9881381}}

@article {yasa,
article_type = {journal},
title = {An open-source, high-performance tool for automated sleep staging},
author = {Vallat, Raphael and Walker, Matthew P},
editor = {Peyrache, Adrien and Büchel, Christian and Bagur, Sophie},
volume = 10,
year = 2021,
pub_date = {2021-10-14},
pages = {e70092},
doi = {10.7554/eLife.70092},
journal = {eLife},
}

@article{mne,
  author = {Gramfort, A. and Luessi, M. and Larson, E. and Engemann, D. A. and Strohmeier, D. and Brodbeck, C. and Goj, R. and Jas, M. and Brooks, T. and Parkkonen, L. and Hämäläinen, M.},
  title = {MEG and EEG data analysis with MNE-Python},
  journal = {Frontiers in Neuroscience},
  volume = {7},
  pages = {267},
  year = {2013},
  doi = {10.3389/fnins.2013.00267}
}

@article{eeglab,
  author = {Delorme, Arnaud and Makeig, Scott},
  title = {EEGLAB: an open source toolbox for analysis of single-trial EEG dynamics including independent component analysis},
  journal = {Journal of Neuroscience Methods},
  volume = {134},
  number = {1},
  pages = {9--21},
  year = {2004},
  doi = {10.1016/j.jneumeth.2003.10.009},
  url = {http://www.sccn.ucsd.edu/eeglab/}
}

@article{Chuang2025,
  title = {Augmenting brain-computer interfaces with ART: An artifact removal transformer for reconstructing multichannel EEG signals},
  author = {Chuang, Chun-Hsiang and Chang, Kong-Yi and Huang, Chih-Sheng and Bessas, Anne-Mei},
  journal = {NeuroImage},
  volume = {310},
  pages = {121--123},
  year = {2025},
  publisher = {Elsevier},
  doi = {10.1016/j.neuroimage.2025.121123},
  url = {https://linkinghub.elsevier.com/retrieve/pii/S1053811925001259},
  language = {en}
}

@article{Cai2025,
  title = {DHCT-GAN: Improving EEG Signal Quality with a Dual-Branch Hybrid CNN--Transformer Network},
  author = {Cai, Yinan and Meng, Zhao and Huang, Dian},
  journal = {Sensors},
  volume = {25},
  number = {1},
  pages = {231},
  year = {2025},
  publisher = {MDPI},
  doi = {10.3390/s25010231},
  url = {https://www.mdpi.com/1424-8220/25/1/231},
  language = {en},
  copyright = {Creative Commons Attribution (CC BY) license}
}

@article{Stigt2023,
  title = {The effect of artifact rejection on the performance of a convolutional neural network based algorithm for binary EEG data classification},
  author = {Van Stigt, M. N. and Ruiz Camps, C. and Coutinho, J. M. and Marquering, H. A. and Doelkahar, B. S. and Potters, W. V.},
  journal = {Biomedical Signal Processing and Control},
  volume = {85},
  pages = {105032},
  year = {2023},
  publisher = {Elsevier},
  doi = {10.1016/j.bspc.2023.105032},
}

@article{lorens2025,
      title={EEG Artifact Detection and Correction with Deep Autoencoders}, 
      author={David Aquilué-Llorens and Aureli Soria-Frisch},
      year={2025},
      journal={arXiv},
      url={https://arxiv.org/abs/2502.08686}, 
}

@article{Yu2024,
  title = {A Learnable and Explainable Wavelet Neural Network for EEG Artifacts Detection and Classification},
  author = {Yu, Yifei and Li, Yuanxiang and Zhou, Yunqing and Wang, Yingyan and Wang, Jiwen},
  journal = {IEEE Transactions on Neural Systems and Rehabilitation Engineering},
  volume = {32},
  pages = {3358--3368},
  year = {2024},
  publisher = {IEEE},
  doi = {10.1109/TNSRE.2024.10659751},
}

@article{Seeuws2024,
  title = {Avoiding Post-Processing With Event-Based Detection in Biomedical Signals},
  author = {Seeuws, Nick and De Vos, Maarten and Bertrand, Alexander},
  journal = {IEEE Transactions on Biomedical Engineering},
  volume = {71},
  number = {8},
  pages = {2442--2453},
  year = {2024},
  publisher = {IEEE},
  doi = {10.1109/TBME.2024.10465660},
}

@article{yadav2025,
title = {Evaluating SMOTE-ENN impact on EEG signal seizure detection and classification},
journal = {Procedia Computer Science},
volume = {258},
pages = {2920-2929},
year = {2025},
note = {International Conference on Machine Learning and Data Engineering},
doi = {https://doi.org/10.1016/j.procs.2025.04.552},
author = {Amit Kumar Yadav and Bashir Alam and Om Pal},
}

@article{Wang2024,
  author    = {Wang, Chenlong and Liu, Lei and Zhuo, Wenhai and Xie, Yun},
  title     = {An Epileptic {EEG} Detection Method Based on Data Augmentation and Lightweight Neural Network},
  journal   = {IEEE Journal of Translational Engineering in Health and Medicine},
  year      = {2024},
  volume    = {12},
  pages     = {22--31},
  doi       = {10.1109/JTEHM.2023.3308196},
}

@article{Malafeev2019,
author = {Malafeev, Alexander and Omlin, Ximena and Wierzbicka, Aleksandra and Wichniak, Adam and Jernajczyk, Wojciech and Riener, Robert and Achermann, Peter},
title = {Automatic artefact detection in single-channel sleep EEG recordings},
journal = {Journal of Sleep Research},
volume = {28},
number = {2},
pages = {e12679},
doi = {https://doi.org/10.1111/jsr.12679},
url = {https://onlinelibrary.wiley.com/doi/abs/10.1111/jsr.12679},
year = {2019}
}
\begin{appendices}
\newpage
\section[Appendix]{Qualitative validation of SMOTE artifact epochs.}
\label{appendix:A}

\begin{figure}[h]
    \centering
    \includegraphics[width=0.99\textwidth]{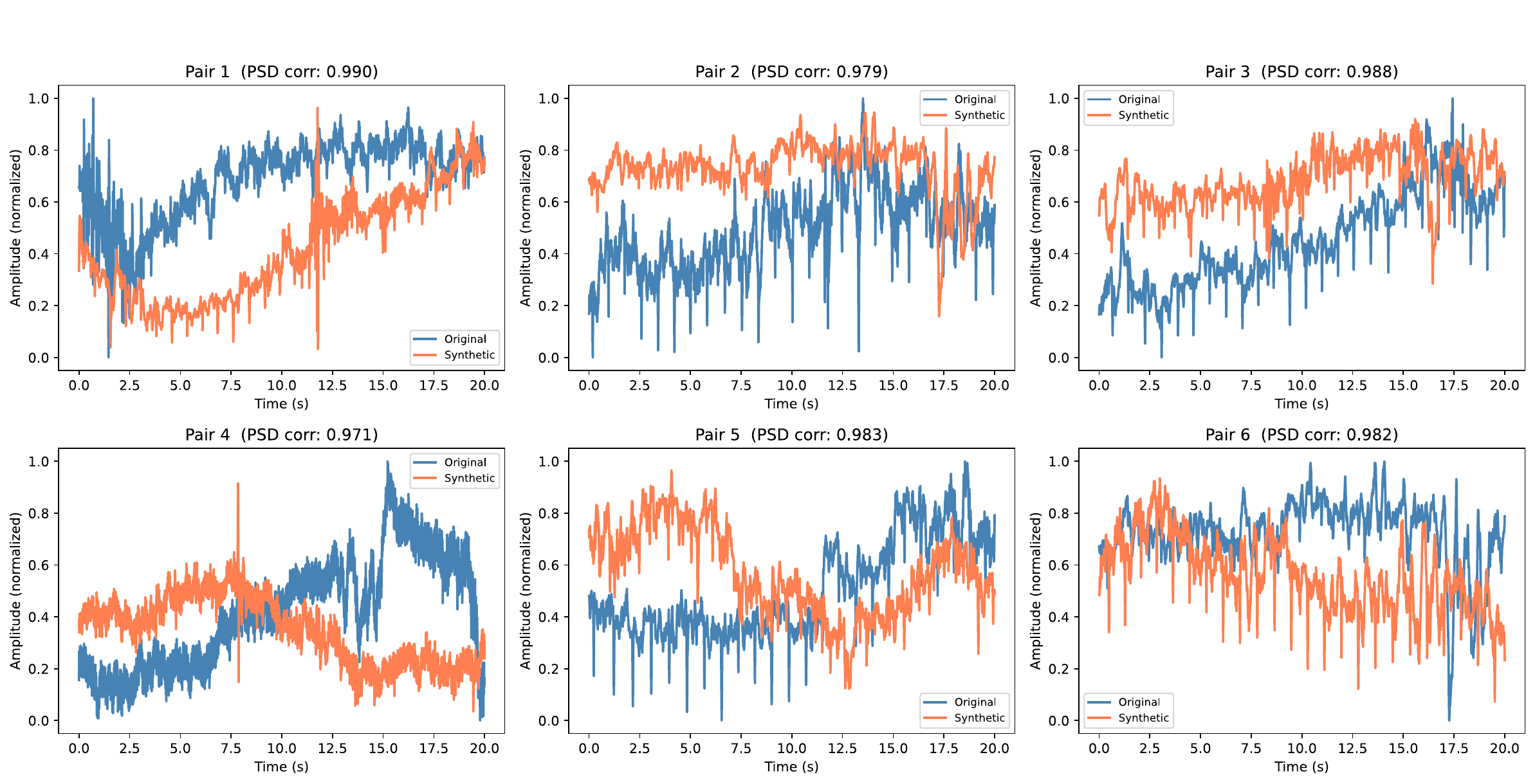}
    \caption{Representative PSD-matched pairs of original (blue) and SMOTE-generated synthetic (red) 20-second artifact epochs. Each subplot shows a randomly selected pair matched by power spectral density (PSD) characteristics. All signals are min-max normalized to [0, 1].}
    \label{fig:appendix_smote}
\end{figure}

\begin{table}[h]
\centering
\caption{Spectral properties of PSD-matched original and synthetic artifact pairs.}
\label{tab:appendix_psd}
\begin{tabular}{cccccccc}
\hline
\thead{\textbf{Pair}} & \thead{\textbf{Type}} & \thead{\textbf{Dominant} \\ \textbf{freq. (Hz)}} & \thead{\textbf{Spectral} \\ \textbf{centroid (Hz)}} & \thead{\textbf{Delta} \\ \textbf{(0.5--4 Hz)}} & \thead{\textbf{Theta} \\ \textbf{(4--8 Hz)}} & \thead{\textbf{Alpha} \\ \textbf{(8--13 Hz)}} & \thead{\textbf{Beta} \\ \textbf{(13--30 Hz)}} \\
\hline
\multirow{2}{*}{1} & Original   & 0.49 & 9.39 & 0.0011 & 0.0004 & 0.0004 & 0.0010 \\
                   & Synthetic  & 0.49 & 8.42 & 0.0008 & 0.0002 & 0.0002 & 0.0006 \\
\hline
\multirow{2}{*}{2} & Original   & 1.46 & 5.33 & 0.0051 & 0.0009 & 0.0006 & 0.0007 \\
                   & Synthetic  & 1.46 & 2.31 & 0.0030 & 0.0003 & 0.0001 & 0.0001 \\
\hline
\multirow{2}{*}{3} & Original   & 0.98 & 5.02 & 0.0015 & 0.0005 & 0.0003 & 0.0004 \\
                   & Synthetic  & 0.49 & 4.10 & 0.0017 & 0.0005 & 0.0003 & 0.0002 \\
\hline
\multirow{2}{*}{4} & Original   & 0.49 & 17.52 & 0.0013 & 0.0001 & 0.0002 & 0.0001 \\
                   & Synthetic  & 49.80 & 17.90 & 0.0005 & 0.0002 & 0.0002 & 0.0001 \\
\hline
\multirow{2}{*}{5} & Original   & 0.49 & 4.67 & 0.0020 & 0.0007 & 0.0005 & 0.0006 \\
                   & Synthetic  & 0.49 & 4.38 & 0.0016 & 0.0006 & 0.0004 & 0.0003 \\
\hline
\multirow{2}{*}{6} & Original   & 1.46 & 2.22 & 0.0058 & 0.0005 & 0.0002 & 0.0001 \\
                   & Synthetic  & 1.46 & 4.94 & 0.0036 & 0.0008 & 0.0005 & 0.0006 \\
\hline
\end{tabular}
\end{table}
\end{appendices}
\end{document}